\documentclass{PoS}
\usepackage{natbib}

\title{A catalog of narrow-line Seyfert 1 galaxies in the southern hemisphere}
\ShortTitle{A NLS1 catalog in the southern sky}

\author{
\speaker{S. Chen} $^{1,2,3}$ \thanks{I would like to thank the organizers, for having realized a pleasant and scientifically fruitful conference, and giving me an opportunity to present my work.}, M. Berton $^{4,5}$, G. La Mura $^{1}$, E. Congiu $^{1,6}$, V. Cracco $^{1}$, L. Foschini $^{6}$, J.H. Fan $^{2}$, S. Ciroi $^{1,7}$, P. Rafanelli $^{1}$, and D. Bastieri $^{1,3}$. \\
\llap{$^1$} Dipartimento di Fisica e Astronomia "G. Galilei", Universit\`a di Padova, Padova, Italy \\
\llap{$^2$} Center for Astrophysics, Guangzhou University, 510006, Guangzhou, China \\
\llap{$^3$} Istituto Nazionale di Fisica Nucleare (INFN), Sezione di Padova, 35131, Padova, Italy \\
\llap{$^4$} Finnish Centre for Astronomy with ESO (FINCA), University of Turku, Quantum, Vesilinnantie 5, 20014 University of Turku, Finland \\
\llap{$^5$} Aalto University Mets\"ahovi Radio Observatory, Mets\"ahovintie 114, FIN-02540 Kylm\"al\"a, Finland \\
\llap{$^6$} INAF - Osservatorio Astronomico di Brera, Via E. Bianchi 46, 23807, Merate (LC), Italy \\
\llap{$^7$} INAF - Osservatorio Astronomico di Padova, Vicolo dell'osservatorio 5, 35122, Padova, Italy \\

E-mail: \email{sina.chen@phd.unipd.it}
}

\abstract{We present a new accurate sample of narrow-line Seyfert 1 galaxies (NLS1s) in the southern hemisphere from the Six-degree Field Galaxy Survey (6dFGS). Based on the optical spectral features, 167 sources were classified as NLS1s. We derived flux-calibrated spectra in the sample for the first time. Strong luminosity correlations between the continuum and the emission lines were found. We estimated their central black hole masses and Eddington ratios, which are lying in a typical range of NLS1s. In the sample, 23 NLS1s were detected at radio frequencies and 12 of them are radio-loud. We publish the X-ray data analysis of a campaign of observations carried out by the \textit{Swift} X-ray Telescope.}

\FullConference{Revisiting narrow-line Seyfert 1 galaxies and their place in the Universe - NLS1 Padova \\
9-13 April 2018 \\
Padova Botanical Garden, Italy}

\begin{document}

\section{Introduction}

Narrow-line Seyfert 1 galaxies (NLS1s) are a peculiar subclass of active galactic nuclei (AGN). The classification is based on their optical spectral properties, having a full width at half maximum FWHM (H$\beta$) $<$ 2000 km s$^{-1}$, and a flux ratio of [O III]$\lambda$5007 / H$\beta$ $<$ 3 \citep{Osterbrock1985, 1989ApJ...342..224G}. Most of their optical spectra show strong Fe II multiplets emission, which is a sign that the broad-line region (BLR) and the accretion disk are visible \citep{2008MNRAS.385...53M, 2011nlsg.confE...2P}. It is believed that NLS1s host a relatively under-massive central black hole compared to broad-line Seyfert 1 galaxies (BLS1s), typically in the range of $M_{BH} \sim 10^{6-8} M_{\odot}$ for NLS1s and $ 10^{7-8} M_{\odot}$ for BLS1s \citep{2000MNRAS.314L..17M}. This suggests that NLS1s might be a young and fast-growing phase of AGN \citep{2000MNRAS.314L..17M, 2000NewAR..44..455G}. However, our current knowledge is too poor yet to reach a conclusion.

To better understand the nature of NLS1s, observations with advanced facilities will be necessary. The aim of this work is to create a new NLS1 sample that can be observed by large telescopes located in the southern hemisphere, for the purpose of investigating the peculiarity of NLS1s with respect to other classes of AGN. The full version of this study has been published by \citet{Chen2018}.

\section{A NLS1 sample in the southern hemisphere} 

We exploited the large unexplored archive of the third and final data release for the Six-degree Field Galaxy Survey (6dFGS DR3) \citep{Jones2009}. Optical spectra were first brought to the rest frame. FWHM of H$\beta$ and [O III] lines, and flux ratio of [O III] / H$\beta$ were calculated.

According to the criteria of 600 km s$^{-1}$ $<$ FWHM(H$\beta$) $<$ 2200 km s$^{-1}$ \citep{2012MNRAS.427.1266V, Cracco2016} and flux ratio of [O III] / H$\beta$ $<$ 3 \citep{Osterbrock1985}, as well as considering the visibility of Fe II multiplets and the H$\beta$ line profile \citep{2013A&A...549A.100K} in the optical spectra, we created a new accurate sample of 167 NLS1s from the 6dFGS in the southern hemisphere.

\subsection{Flux calibration}

Since the 6dFGS does not provide flux information on the spectra, we derived a flux calibration for these 167 NLS1s identified in our sample. To evaluate the reliability of the calibration, we compared the relation between the continuum flux at 5100 $\mathring{A}$ and the optical B-band magnitude of our 6dFGS re-calibrated spectra with that holding for a sample of 296 flux calibrated spectra, selected and investigated by \citet{Cracco2016} from the Sloan Digital Sky Survey Data Release 7 (SDSS DR7) \citep{Abazajian2009}. The two relations are in good agreement.

An additional test for our calibration was carried out. We observed 15 objects in the 6dFGS sample to obtain their optical spectra using the Asiago Astrophysical Observatory (Italy) 1.22 m and 1.82 m telescopes in October and November 2017. We compared the flux-calibrated spectra obtained from the 6dFGS and the Asiago telescopes, confirming that the fluxes in the H$\beta$, [O III], and continuum regions are consistent generally, even though the uncertainties on the blue and red extremes of the spectral range are not negligible. To estimate the uncertainty of our flux calibration, we computed the flux differences of [O III] line between the 6dFGS and Asiago spectra, finding an average uncertainty of 38.7$\%$. Based on these properties, we suppose that for our purpose the flux calibration result is acceptable, though introducing some degree of uncertainty.

\subsection{Luminosity correlation}

We first subtracted the host galaxy contribution from our flux-calibrated spectra, in order to obtain an estimate of the light originally emitted by the AGN alone, following the procedure described in \citet{LaMura2007}. After the host galaxy correction, the monochromatic luminosity at 5100 $\mathring{A}$ was calculated. To estimate the flux and luminosity of H$\beta$ and [O III] lines, we subtracted the continuum and Fe II multiplets \citep{2010ApJS..189...15K} from the flux-calibrated spectra. The flux and luminosity of both emission lines were calculated by integrating the H$\beta$ and [O III] line profiles and assuming an isotropic radiation.

Strong correlations of L(H$\beta$) - $\lambda$L$_{\lambda}$(5100$\mathring{A}$) and L([O III]) - $\lambda$L$_{\lambda}$(5100$\mathring{A}$) were obtained using the linear regression with a Bayesian method \citep{Kelly2007}, which are in agreement with various works in literature \citep{2005ApJ...630..122G, 2006ApJS..166..128Z, 2010ApJ...723..409G}.

\subsection{Black hole mass and Eddington ratio}

We estimated the central black hole mass following the method in \citet{Foschini2015} and \citet{Berton2015}. For each NLS1 galaxy in our sample, the virial mass of the central black hole is defined as
\begin{equation}
M_{BH} = f \left( \frac{R_{BLR} \sigma^{2}_{line}}{G} \right),
\end{equation}
where $R_{BLR}$ is the size of the BLR measured by reverberation \citep{Bentz2013}, $\sigma_{line}$ is the line dispersion determined from the H$\beta$ broad components, $G$ is the gravitational constant, and $f$ is a dimensionless scale factor \citep{Grier2013}. The Eddington ratio is defined as $\epsilon = L_{bol} / L_{Edd}$, where the Eddington luminosity is
\begin{equation}
L_{Edd} = 1.3 \times 10^{38} \left(\frac{M_{BH}}{M_{\odot}}\right) \, \rm{erg \, s^{-1}},
\end{equation}
and the bolometric luminosity is estimated assuming $L_{bol} = 9 \lambda L_{\lambda} (5100\mathring{A})$ \citep{2000ApJ...533..631K}.

The masses of the central black hole $M_{BH}$ range from $8.1 \times 10^{5} M_{\odot}$ to $7.8 \times 10^{7} M_{\odot}$ with a median value of $8.6 \times 10^{6} M_{\odot}$. The Eddington ratios $L_{bol} / L_{Edd}$ span between 0.07 to 5.35 with a median value of 0.96. Our result is in agreement with a recent study by \citet{2017ApJS..229...39R}, who analyzed a new sample that contains NLS1s and BLS1s from the SDSS DR12. They found that the black hole masses have an average of $\log M_{BH}$ of $6.9 M_{\odot}$ for NLS1s and $8.0 M_{\odot}$ for BLS1s, the Eddingtion ratios have an average of $\log \epsilon$ of -0.34 for NLS1s and -1.03 for BLS1s. This confirms that NLS1s have lower black hole mass and higher Eddington ratio than BLS1s.

\subsection{Radio sources}

We subsequently cross-matched these 167 NLS1s with several radio surveys covering the southern hemisphere, including the NVSS \citep{1998AJ....115.1693C}, the SUMSS \citep{2003MNRAS.342.1117M}, and the AT20G \citep{2010MNRAS.402.2403M}, within a search radius of 5 arcsec. Totally 23 (13.8$\%$) sources have an associated radio counterpart. We calculated the radio loudness $R_{L} = F_{5 GHz} / F_{4400 \mathring{A}}$ \citep{1989AJ.....98.1195K} for each object, and found 12 (7.0$\%$ of the whole sample) radio-loud (RL, $R_{L} > 10$) NLS1s and 11 radio-quiet (RQ, $R_{L} < 10$) NLS1s.

The two-sample Kolmogorov-Smirnov (K-S) test was used to examine whether the parent population of the RL and RQ subsample is the same. We applied the null hypothesis of two distributions originated from the same population of sources, and the rejection of the null hypothesis at a 95$\%$ confidence level corresponding to a value of $p \leq 0.05$. The K-S tests of redshift ($p = 1.8 \times 10^{-3}$) and radio luminosity ($p = 2.0 \times 10^{-4}$) suggest that RL and RQ NLS1s in the sample have different populations. However, the K-S tests of black hole mass ($p = 0.47$) and optical B-band luminosity ($p = 0.055$) argue that these RL and RQ sources have the same origin. The cumulative distributions of radio sample shows as well that RL sources tend to have higher redshift, more massive black hole, higher radio and optical luminosity than RQ sources on average.

\section{X-rays}

With the aim of further investigating the X-rays properties of radio-emitting NLS1s in the 6dFGS sample, we carried out an observational campaign with the \textit{Swift} X-Ray Telescope (XRT) \citep{Burrows2005} from November 2017 to March 2018, details are listed in Table~\ref{swift}. These objects are strong candidates of harboring relativistic jet activity. The analysis of X-ray data might contribute to clarify the role of central black hole, because it directly probes the nature and intensity of the ionizing radiation continuum.

\begin{table}[ht]
\caption{Details of the \textit{Swift} XRT observations.}
\label{swift}
\resizebox{\textwidth}{!}{
\centering
\begin{tabular}{cccccc}
\hline
\hline
Name & RA & Dec & z & cps & exp \\
- & (hh mm ss) & (dd mm ss) & - & (10$^{-2}$ ct \, s$^{-1}$) & (sec) \\
\hline
6dFGS gJ012237.5-264646	& 01 22 37.52 & -26 46 45.9 & 0.417 & 2.99 & 2308 \\
6dFGS gJ042256.6-185442 & 04 22 56.56 & -18 54 42.3 & 0.064 & 0.19 & 13020 \\
6dFGS gJ084628.7-121409 & 08 46 28.67 & -12 14 09.3 & 0.108 & 17.73 & 3943 \\
6dFGS gJ104208.9-400032 & 10 42 08.92 & -40 00 31.6 & 0.386 & 1.04 & 15810 \\
6dFGS gJ114738.9-214508 & 11 47 38.87 & -21 45 07.7 & 0.219 & 0.19 & 7827 \\
6dFGS gJ143438.4-425405 & 14 34 38.43 & -42 54 05.3 & 0.115 & 7.65 & 5309 \\
6dFGS gJ150012.8-724840 & 15 00 12.81 & -72 48 40.3 & 0.141 & 3.87 & 8626 \\
6dFGS gJ151222.5-333334 & 15 12 22.47 & -33 33 34.2 & 0.023 & 0.16 & 7347 \\
6dFGS gJ151515.2-782012 & 15 15 15.20 & -78 20 12.0 & 0.259 & 4.46 & 3878 \\
6dFGS gJ152228.7-064441 & 15 22 28.72 & -06 44 40.9 & 0.083 & 19.86 & 3681 \\
6dFGS gJ164610.4-112404 & 16 46 10.39 & -11 24 04.2 & 0.074 & 31.90 & 5549 \\
6dFGS gJ205920.7-314735 & 20 59 20.74 & -31 47 34.8 & 0.073 & 8.28 & 6548 \\
\hline
\end{tabular}}
\textbf{Notes.} Column (1) Name in the 6dFGS. (2) Right Ascension. (3) Declination. (4) Redshift. (5) Count rate. (6) Exposure time.
\end{table}

Data analysis was performed following standard procedures (XRTPIPELINE). We considered the data observed in photon counting (PC) mode. The channels with discarded data \footnote{Some are below the lower discriminator of the instrument and therefore do not contain valid data. Some have imperfect background subtraction at the margins of the pass band. Some may not contain enough counts for $\chi^2$ to be strictly meaningful.} and energies out of the broad-band range provided by the \textit{Swift} XRT detector (below 0.3 keV and above 10.0 keV) were excluded from the fit. The XRT spectral counts were rebinned to have 20 counts per bin in order to apply the Chi-squared $\chi^2$ statistics. When the number of counts was not enough to do this, we applied the unbinned likelihood and forced to use the Cash statistics \citep{Cash1979}. Each spectrum was fitted with an ISM grain absorption (tbabs) and a redshifted power-law (zpowerlaw) model using XSPEC \footnote{https://heasarc.nasa.gov/xanadu/xspec/manual/node1.html.}. The \textit{Swift} XRT spectral results are reported in Table~\ref{x-ray}.

\begin{table}[ht]
\caption{X-ray properties of the radio sample.}
\label{x-ray}
\resizebox{\textwidth}{!}{
\centering
\begin{tabular}{ccccccc}
\hline
\hline
Name & $N_{H}$ & $\Gamma$ & F$_{0.3-10.0 \, keV}$ & L$_{0.3-10.0 \, keV}$ & Stat. & Val./d.o.f. \\
- & (10$^{20}$ cm$^{-2}$) & - & (10$^{-12}$ erg \, cm$^{-2}$ s$^{-1}$) & (10$^{43}$ erg \, s$^{-1}$) & - & - \\
\hline
6dFGS gJ012237.5-264646 & 1.21 & 2.52 $\pm$ 0.4 & 0.80 & 56.13 & C & 36.13/44 \\
6dFGS gJ042256.6-185442 & 3.18 & 0.74 $\pm$ 0.5 & 0.20 & 0.18 & C & 44.62/20 \\
6dFGS gJ084628.7-121409 & 6.04 & 2.81 $\pm$ 0.3 & 6.09 & 18.52 & $\chi^2$ & 25.82/27 \\
6dFGS gJ104208.9-400032 & 6.67 & 2.61 $\pm$ 0.3 & 0.39 & 20.34 & $\chi^2$ & 3.10/5 \\
6dFGS gJ114738.9-214508 & 3.69 & 0.90 $\pm$ 0.8 & 0.12 & 1.33 & C & 13.68/9 \\
6dFGS gJ143438.4-425405 & 5.38 & 1.74 $\pm$ 0.3 & 3.07 & 10.04 & $\chi^2$ & 13.72/15 \\
6dFGS gJ150012.8-724840 & 10.01 & 1.83 $\pm$ 0.3 & 1.56 & 7.88 & $\chi^2$ & 9.03/12 \\
6dFGS gJ151222.5-333334 & 12.60 & 1.08 $\pm$ 0.9 & 0.11 & 0.01 & C & 10.24/8 \\
6dFGS gJ151515.2-782012 & 7.09 & 1.87 $\pm$ 0.3 & 1.63 & 30.97 & $\chi^2$ & 5.70/5 \\
6dFGS gJ152228.7-064441 & 5.89 & 2.59 $\pm$ 0.1 & 6.70 & 11.50 & $\chi^2$ & 19.08/29 \\
6dFGS gJ164610.4-112404 & 10.22 & 1.59 $\pm$ 0.1 & 15.55 & 19.99 & $\chi^2$ & 91.79/73 \\
6dFGS gJ205920.7-314735 & 9.38 & 2.31 $\pm$ 0.3 & 2.71 & 3.49 & $\chi^2$ & 13.35/21 \\
\hline
\end{tabular}}
\textbf{Notes.} Column (1) Name in the 6dFGS. (2) Column density. (3) Photon index. (4) Flux in 0.3 - 10.0 keV energy range. (5) Luminosity in 0.3 - 10.0 keV energy range. (6) $\chi^2$: Chi-squared statistics, C: Cash statistics. (7) Value / degrees of freedom.
\end{table}

\begin{figure}[ht]
\centering
\includegraphics[width=0.8\textwidth]{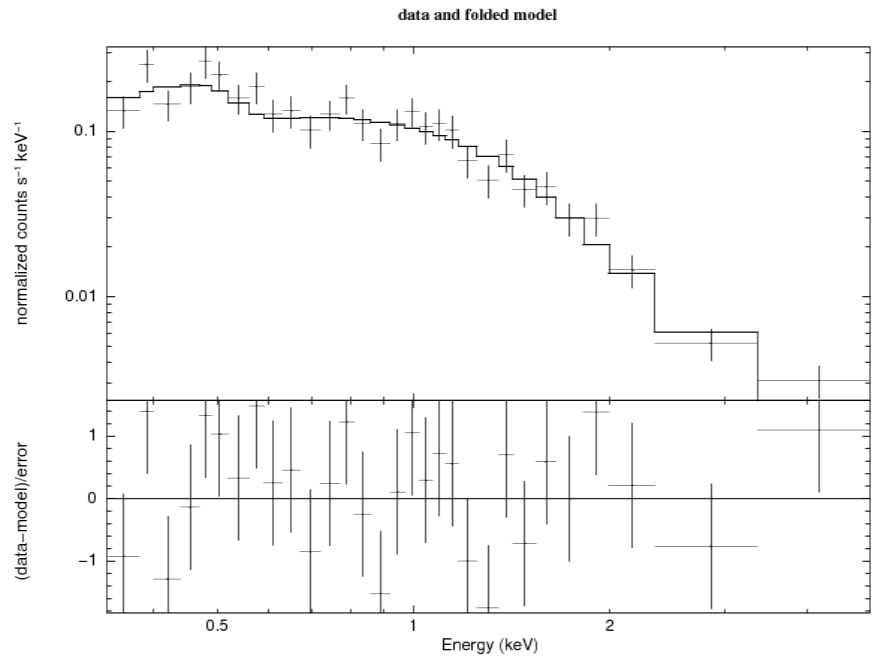}
\caption{The \textit{Swift} XRT spectrum of 6dFGS gJ084628.7-121409 fitted with a galactic absorption and a redshifted power-law model.}
\label{xspec}
\end{figure}

The average X-ray spectral index in the 0.3 - 10.0 keV energy band is $\alpha$ = 0.88 $\pm$ 0.4 (power-law model, $\Gamma = \alpha + 1$). Compared to the mean values of flat-spectrum radio quasars (FSRQs) ($\alpha$ = 0.58), BL Lacertae objects (BL Lac) ($\alpha$ = 1.3), and BLS1s ($\alpha$ = 1.1), the average X-ray spectral index of NLS1s is softer than FSRQs while harder than BL Lac and BLS1s \citep{Foschini2015}. An example of the \textit{Swift} XRT spectrum is shown in Fig.~\ref{xspec}. However, it is hard to make any conclusion due to the limited number of observed objects and photon counts from the \textit{Swift} observations.

\section{Summary}

In this work, we selected a sample of 167 NLS1s from the 6dFGS increasing the number of known NLS1s in the southern hemisphere. Flux calibration for the optical spectra in the sample was derived with reliable results. The luminosity correlations of L(H$\beta$) - $\lambda$L$_{\lambda}$(5100$\mathring{A}$) and L([O III]) - $\lambda$L$_{\lambda}$(5100$\mathring{A}$) were confirmed. The mass of central black holes are in the range of $M_{BH} \sim 10^{5.9-7.9} M_{\odot}$ and confirm that NLS1s have lower black hole mass than BLS1s. We found that 23 NLS1s in the sample have associated radio counterparts, including 12 RL and 11 RQ. RL sources tend to have higher redshift, a more massive black hole, and higher radio and optical luminosities than RQ sources.

In addition, we carried out an observational campaign of radio-emitting NLS1s in the 6dFGS sample with the \textit{Swift} XRT. The X-ray spectra were fitted with a galactic absorption and a redshifted power-law models in the 0.3 - 10.0 keV energy range. The average spectral index of NLS1s is softer than FSRQs while harder than BL Lac and BLS1s.

Finally, we remark that these results should be taken with caution because the number of NLS1s in the 6dFGS sample, in particular sources having radio and X-ray counterpart, is still limited. Further researches with larger samples, higher resolution and sensitivity observations will be necessary to understand the physical mechanism and evolution of NLS1s, with respect to other types of AGN.

\section*{Acknowledgements}

JHFan's work is partially supported by the National Natural Science Foundation of China (NSFC 11733001 and NSFC U1531245). This conference has been organized with the support of the Department of Physics and Astronomy ''Galileo Galilei'', the University of Padova, the National Institute of Astrophysics INAF, the Padova Planetarium, and the RadioNet consortium. RadioNet has received funding from the European Union's Horizon 2020 research and innovation programme under grant agreement No~730562. 

\bibliographystyle{rusnat}
\bibliography{bibliography}

\end{document}